\documentclass[preprint, review, 12pt]{elsarticle}

\usepackage{setspace}
\usepackage{xcolor}
\usepackage{graphicx}
\usepackage[labelfont=bf,labelsep=period,skip = 0pt]{caption}
\usepackage{xr-hyper}
\usepackage{chemformula} 
\usepackage[T1]{fontenc} 
\usepackage{multirow}
\usepackage{comment}
\usepackage[version=3]{mhchem}

\usepackage{booktabs}
\usepackage{xr}

\biboptions{numbers,sort&compress}
\journal{Fuel}

\begin{document}
\begin{doublespace}
\begin{frontmatter}

\title{Enhanced Accumulation of Bitumen Residue in a Highly Concentrated Tailings Flow by Microbubbles from In-situ Catalytic Decomposition of Hydrogen Peroxide}

\author[First]{Kaiyu Zhou}
\address[First]{Department of Chemical and Materials Engineering, University of Alberta, Alberta T6G 1H9, Canada}
\author[First]{Somasekhara Goud Sontti}
\author[Second]{Joe Zhou}
\address[Second]{Disruptive Separation Technology Ltd.(DSTL), Edmonton, Alberta T6X 1M5, Canada}
\author[First]{Xuehua Zhang\corref{cor}}
\ead{xuehua.zhang@ualberta.ca}

\cortext[cor]{Corresponding author}

\begin{abstract}
    
The massive volume of oil sands tailings has been one of the most challenging environmental issues. In this work, we experimentally explore a simple and effective approach to bitumen residue separation from a highly concentrated slurry flow of the artificial oil sands tailings. By utilizing microbubbles from in-situ catalytic decomposition of hydrogen peroxide ($H_{2}O_{2}$) at low concentrations,  bitumen aggregation is enhanced on the top part of the hydrotransport pipeline. 
The microscopic image analysis revealed the in-situ formation of microbubbles and confirmed that magnetic particles present in the slurries contributed to the fast release of the gas products and bubble formation from $H_{2}O_{2}$ decomposition.    A high-speed camera was applied to capture images of the tailings flow in the pipeline through a transparent view window. A large number of tiny bubbles were identified post to the injection of $H_{2}O_{2}$ to the slurry flow.  More than 70 \% bitumen could be recovered from a lab-scale pipeline loop within 30 mins after $H_{2}O_{2}$ injection.
The bitumen recovery efficiency from the collected froth was quantitatively compared under seven conditions with varied dosages, the concentration of $H_{2}O_{2}$, and the amount of magnetic solids in the slurries. Our results confirmed  that the total dosage of $H_{2}O_{2}$ is the dominant factor in in-situ microbubble formation for enhanced bitumen aggregation in the flow. Importantly, microbubbles were generated rapidly in the real mature fine tailings. The results from our study provide insights into the preferential distribution of oil residue in the flow during hydrotransport without the requirement for an additional device. Removal of oily residues from concentrated slurries may bring economical and environmental advantages. 
\end{abstract}


\end{frontmatter}

\section{Introduction}
Oil sands tailings are composed of process water, sand, silt, clay, and unrecovered hydrocarbons as well as minerals of great value \cite{small2015emissions,chalaturnyk2002management}. The discharged tailings from primary separation cell are transported to the tailing ponds by pipeline \cite{sadeghi2022computational}. Organic waste due to incomplete bitumen recovery is continuously accumulated in the tailing ponds. It generates hazardous emissions and endangers the waterfowl. Therefore, the removal of residual bitumen is an impending matter in order to meet environmental requirements and protect the shelter for wildlife \cite{da2021recovery}. Basically, there are several methods for separating bitumen from oil sand ores, such as solvent extraction, thermal treatment, and bubble flotation \cite{li2011ionic,mukhametshina2014evaluation}.
By contrast, bubble flotation is energy-efficient and environmentally friendly, which is also applied for the extraction of various ores including zinc, copper, as well as coal \cite{dai1999particle,you2017investigation}.

Many efforts have been made to investigate the role of bubbles in enhancing the efficiency of flotation and separation. Previous authors find that microbubbles show a better performance in contact with hydrophobic substances compared to large bubbles \cite{zhou2020role,wang2020regulation}. The high surface-to-volume ratio and long residence time of microbubbles increase the probability of a collision. After colliding, microbubbles favor attachment to the dispersed hydrophobic matter and form aggregates that have a higher resistance to detachment by shearing. In a two-stage attachment model, microbubbles frosted on the surface of aggregates can coalesce with flotation-size bubbles, and hence flotation kinetics is accelerated by a higher buoyancy force \cite{tao2005role}.

To generate microbubbles, cavitation induced by pressure fluctuation has widespread applications \cite{zhou2009role}, such as hydrodynamic cavitation and acoustic cavitation \cite{li2019study,leighton1995bubble}.
The previous study used a modified venturi tube to generate microbubbles upstream of the oil sands tailings of 50 wt\% solid contents for separating bitumen residue of 5 wt\% in the pipeline loop \cite{motamed2020microbubble}. This venturi tube generates bubbles by hydrodynamic cavitation and shearing the sucked gas. The result shows that approximately 50 \% bitumen can be recovered with the injection of air microbubbles from a highly concentrated tailing stream in the pipeline \cite{motamed2020microbubble}. Complicatedly, bubble generation by hydrodynamic cavitation is associated with the venturi tube design, the content of dissolved gas, and temperature \cite{cai2009dynamic,zhou2009role}.
Another effective method is the depressurization of supersaturated fluid for microbubble generation \cite{vlyssides2004bubble}. However, it is impractical to apply compressed pressure in pipeline transport in consideration of safety and high energy consumption \cite{zimmerman2008microbubble}. 
In addition, microfluidics devices are able to form a great concentration of uniform microbubbles per unit volume with surfactant \cite{caudwell2022protein,peyman2012expanding}. But scaling up applications to the waste treatment industry is challenging. Also, water electrolysis can generate both hydrogen and oxygen microbubbles in the conductive fluid \cite{matsuura2019control}. 

The addition of devices for microbubble generation in the pipeline adds to the difficulty in construction and operation. It is well known that reaction of chemicals such as $NaHCO_{3}$, $CaCO_{3}$ and $H_{2}O_{2}$ \cite{fu2009bubble,min2015ph}, can generate a large amount of gas bubbles. Compared to the chemicals in the state of powder, $H_{2}O_{2}$ has a more homogeneous dispersion in the oil sands tailings for bubble generation. Moreover, $H_{2}O_{2}$ decomposition is enhanced by the alkaline tailing slurry, and the residual iron oxides \cite{molamahmood2022catalyzed}. On the contrary, bubbles are generated in the acidic fluid by reactions of $NaHCO_{3}$ and $CaCO_{3}$. Last but not least is that the final product of $H_{2}O_{2}$ is water and oxygen, which are relatively stable. The residual $H_{2}O_{2}$ can be continuously decomposed in the tailing ponds.

As illustrated above, bubble size plays an important role in the efficiency of recovery.
The size of bubbles generated by $H_{2}O_{2}$ decomposition is affected by several factors. Huang etal.~
\cite{huang2013catalytic} pointed out that the high concentration of $H_{2}O_{2}$ contributes to larger bubbles as the quick supplement of $H_{2}O_{2}$ on catalytic sites accelerates the growth of bubbles. Furthermore, the surface wettability of catalysts significantly influences the departure diameter and growth period of bubbles \cite{nam2009experimental}. The hydrophobic surface functions as an anchor for growing bubbles and thus contributes to larger bubbles. Nevertheless, the detachment of bubbles is enhanced by shearing in the turbulent tailings flow \cite{abu2021vapour}. In this case, the growth of bubbles might be restrained efficiently. In the study by Okawa et al.~\cite{okawa2011recovery}, it was discovered that stirring a frothing agent, namely $H_{2}O_{2}$, in a vessel could be used to recover bitumen from oil sands ores. However, they found that the efficiency of this method was low, with less than 10 wt$\%$ bitumen being recovered when using $H_{2}O_{2}$ at a concentration of 1000 ppm. Additionally, few studies have investigated the use of $H_{2}O_{2}$ in flotation behavior. 


This work aims to develop a simple and effective approach for enhanced bitumen segregation in highly concentrated oil sands tailings in a lab-scale pipeline loop using $H_{2}O_{2}$. We systematically investigated the combinations of $H_{2}O_{2}$ dosage and injection time on the recovery efficiency of bitumen. Furthermore, we identified the impact of $H_{2}O_{2}$ injection and the effect of magnetic components in tailings on bubble formation and bitumen recovery. In total, seven conditions were compared. To the best of our knowledge, the approach presented in this work is a novel for generating a large number of microbubbles for separation without intensified energy input and additional devices. This study provides new insight into bitumen recovery from concentrate slurry during hydrotransport in a pipeline.

\section{Experimental Section}

\subsection{Preparation of Artificial Oil Sands Tailings}
Artificial oil sands tailings are prepared by following the procedure similar to that in our previous study \cite{zhou2022microbubble}. In brief, sands (Sil Industry Mineral, Edmonton, Canada) are moisturized with 7 \% of the total amount of process water followed by thorough manual blending. Then, the preheated bitumen at 80 \textdegree C is added and mixed with the hydrated sands to be homogeneous by a power agitator for 10 minutes. After that, the rest process water is added to the mixture followed by 2 minutes of blending.

To prepare the artificial oil sands tailings with additional catalysts, the magnetic particles contained in the fresh dry sands were separated by magnetic separation. The separated particles were also applied to understand the effect on the decomposition of $H_{2}O_{2}$. Simultaneously, non-magnetic particles were separated. Further purification was processed until no more magnetic particles were observed. Both separated particles were dried in the oven.
Entrained impurities in the magnetic particles were wiped out using DI water. 
At the beginning of preparation, the magnetic particles were mixed thoroughly with dry sands before addition of process water. The following steps are the same as the sample without additional catalysts.
The properties of artificial oil sands tailings and process water properties are presented in Table \ref{tab1}. In our study, the process water and solids used were identical to those reported in previous literature \cite{motamed2020microbubble,zhou2022microbubble}, so we did not measure the properties of the process water. The manufacturer of the solids used (Sil Industry Mineral, Edmonton, Canada) provided information regarding their composition.


\begin{table}
\centering
\captionsetup{font={normal}}
\renewcommand{\captionfont}{\normalsize}
  \caption{Physical properties of water and artificial tailing composition\cite{zhou2022microbubble}} 
 \label{tab1}
\begin{tabular}{ll}
\toprule 
Parameter  & Value
\\
\hline 
Artificial tailings

\\
Solid ($wt\%$)         &        50
\\
Process water ($wt\%$)        &        49.8
\\
Bitumen ($wt\%$)        &        0.2
\\
Process water properties\cite{motamed2020microbubble}
\\
pH & 7.5
\\
Surface tension ($mN/m$)   &     70.1
\\
Chemical composition & \begin{tabular}[c]{@{}l@{}}Calcium, ammonia, magnesium, sodium, \\ bicarbonate, chloride, sulfate\end{tabular}\\

Solid properties 
\\
Composition ($wt\%$)
\\
\ce{SiO2}   &   93.8
\\
\ce{Al2O3}  &  $<$4.00
\\
\ce{Fe2O3}  &  $<$1.20
\\
\ce{TiO2}  &  $<$0.01
\\
Average size ($\mu$m)  &  74
\\
\hline 
\end{tabular}
\label{propertiesoftailings}
\end{table}

\subsection{Laboratory Hydrotransport Pipeline Loop}
\begin{figure}[ht] 
	\centering
	\includegraphics[width=1\columnwidth]{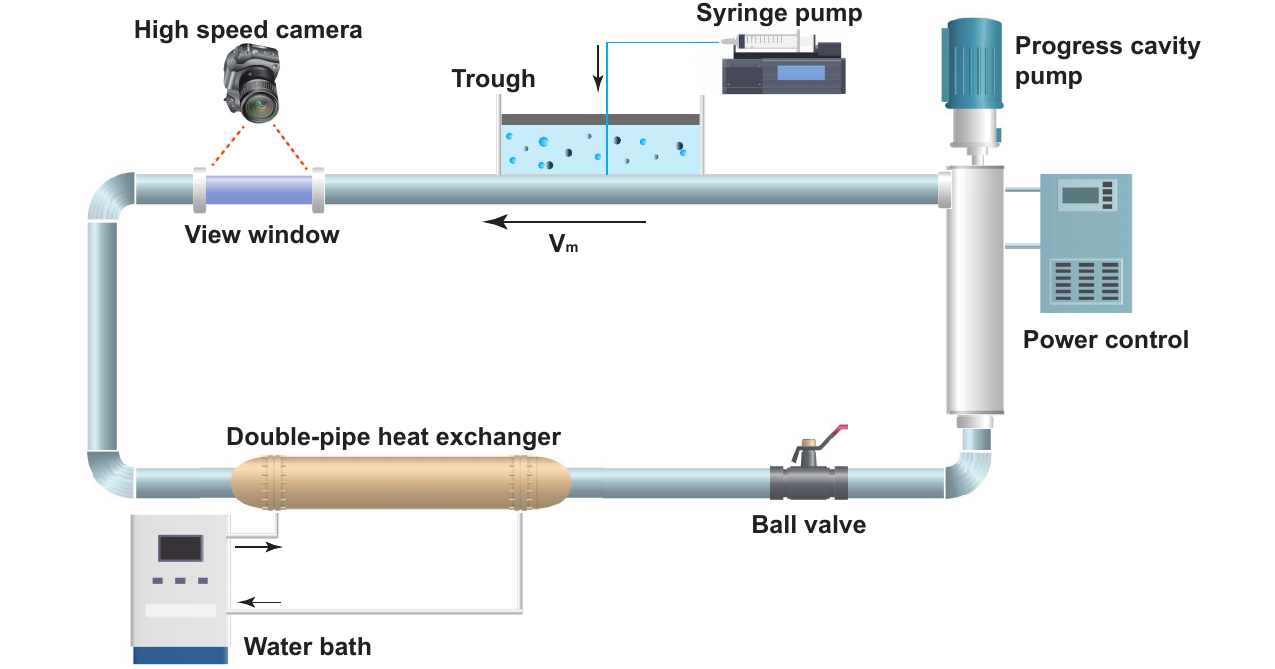} 
    \captionsetup{font={normal}}
    \renewcommand{\captionfont}{\normalsize}
	\caption{Schematic of lab-scale pipeline loop with $H_{2}O_{2}$ injection from the trough; $V_m$: slurry velocity in the main loop \cite{motamed2020microbubble,wallwork2004processibility}.}
	\label{f01}
\end{figure}

A laboratory hydrotransport pipeline loop is designed to emulate the flow behavior of the slurry and operating conditions in the industry. It is applied to process the artificial oil sands tailings to recover bitumen in this study. Figure \ref{f01} shows the Schematic of the hydrotransport pipeline loop composed of a double pipe heat exchanger, progressive cavity pump, trough, and valves.

The pipeline made of stainless steel has an inner diameter of 2.2 $cm$ with 200 $cm$ in length and 68 $cm$ in height. The glass trough at the top allows adding the artificial oil sands tailings and injection of $H_{2}O_{2}$ by a syringe pump. At the beginning, the pipeline was fed with tap water for heating to the set temperature for the purpose of reducing time needed for heating the oil sands tailings. Once the desired temperature is achieved, the tap water was released and the artificial oil sands tailings were introduced to the pipeline. Froths composed of bubbles, solids, process water, and bitumen float to the surface of the trough for collection.

A transparent view window installed at the left of the trough is 40 $cm$ in length with an inner diameter of 17 $mm$. 
The view window facilitates the utilization of high-speed camera which will be discussed in the following section.
At the right part of the loop, it is a progress cavity pump connected to the power control. The flow velocity can be adjusted timely and accurately.

To control the temperature, a double pipe heat exchanger coupled with a programmable circulating bath is installed at the bottom of the pipeline, which allows the water to circulate through the heat exchanger to heat and cool the fluid in the pipeline loop. It takes approximately 20 minutes for the artificial oil sands tailings to reach the set temperature, here 42 \textdegree C. Additionally, a mercury thermometer is inserted into the fluid at the trough for monitoring the temperature. The ball valve near the heat exchanger is used to discharge the waste. After discharge, tap water is injected into the pipeline to wash out the residual solids and bitumen.

\subsection{Froth Collection and Determination of Bitumen Recovery}

In advance to the froth collection, $H_{2}O_{2}$ was injected to generate microbubbles for bitumen recovery. Table \ref{tab2} shows the operating parameters of $H_{2}O_{2}$ injection. The syringe for $H_{2}O_{2}$ was covered with foil in case of decomposition induced by light. The injection rate and duration were controlled by syringe pump. Once the injection was completed, the tube was removed from the trough.

\begin{table}
	\centering
 \captionsetup{font={normal}}
\renewcommand{\captionfont}{\normalsize}
	\caption{Experimental conditions of hydrotransport pipeline loop ($Q$: $H_{2}O_{2}$ dosage, $t$: injection duration, $V_c$: concentration increase rate and $T$: temperature, $V_m$: slurry velocity in the main loop)}
	\label{tab2}
	\begin{tabular}{c|c|c|c|c|c} 
	\hline
		Case & \begin{tabular}[c]{@{}c@{}}$Q$\\ ($mL$)\end{tabular} & \begin{tabular}[c]{@{}c@{}}$t$\\ ($min$)\end{tabular} & \begin{tabular}[c]{@{}c@{}}$V_c$\\~($ppm/min$)\end{tabular} & \begin{tabular}[c]{@{}c@{}}$T$ \\ (\textdegree C)\end{tabular} & \begin{tabular}[c]{@{}c@{}}$V_m$\\ ($m/s$)\end{tabular}  \\ 
		\hline
		\multicolumn{1}{c|}{1}    & \multicolumn{1}{c|}{\multirow{5}{*}{6}}                               & \multicolumn{1}{c|}{2}                                                 & \multicolumn{1}{c|}{250}                                                   & \multirow{7}{*}{42}                              & \multirow{7}{*}{2}                                  \\
		\multicolumn{1}{c|}{2}    & \multicolumn{1}{c|}{}                                                & \multicolumn{1}{c|}{5}                                                 & \multicolumn{1}{c|}{100}                                                    &                                                  &                                                     \\
		\multicolumn{1}{c|}{3}    & \multicolumn{1}{c|}{}                                                 & \multicolumn{1}{c|}{12}                                               & \multicolumn{1}{c|}{42}                                                     &                                                  &                                                     \\
		\multicolumn{1}{c|}{4}    & \multicolumn{1}{c|}{}                                                 & \multicolumn{1}{c|}{18}                                                & \multicolumn{1}{c|}{28}                                                     &                                                  &                                                     \\
		\multicolumn{1}{c|}{5}    & \multicolumn{1}{c|}{}                                                 & \multicolumn{1}{c|}{25}                                                & \multicolumn{1}{c|}{20}                                                     &                                                  &                                                     \\ 
		\cline{2-4}
		\multicolumn{1}{c|}{6}    & \multicolumn{1}{c|}{\multirow{2}{*}{12}}                              & \multicolumn{1}{c|}{5}                                                 & \multicolumn{1}{c|}{200}                                                    &                                                  &                                                     \\
		\multicolumn{1}{c|}{7}    & \multicolumn{1}{c|}{}                                                 & \multicolumn{1}{c|}{25}                                                & \multicolumn{1}{c|}{40}                                                     &                                                  &                                                     \\
		\hline
	\end{tabular}
\end{table}

A spatula is used to collect the recovered froth in the thimbles at four intervals. These thimbles were transferred to the Dean-Stark apparatus to analyze compositions of froth. Bitumen was dissolved in the boiling toluene and water was separated after evaporation and condensation. Thus, only solids were remained in the thimbles. The mass of each recovered composition can be obtained accordingly. The tests under each condition, as shown in Table 2 are conducted three times to ensure the reproducibility. A first-order disappearance kinetic model is used to fit the bitumen recovery with time \cite{zhou2022microbubble}:
\begin{equation}
 R=\text{R}_\infty(1-e^{-kt})  
\end{equation}
where R and $\text{R}_{\infty}$ are designated as the bitumen recovery at time t and infinite, respectively, and k is the bitumen flotation rate constant (min$^{-1}$).

\subsection{Analysis of Bubbles from $H_{2}O_{2}$ Decomposition}
\begin{figure}[!ht] 
	\centering
	\includegraphics[width=0.90\columnwidth]{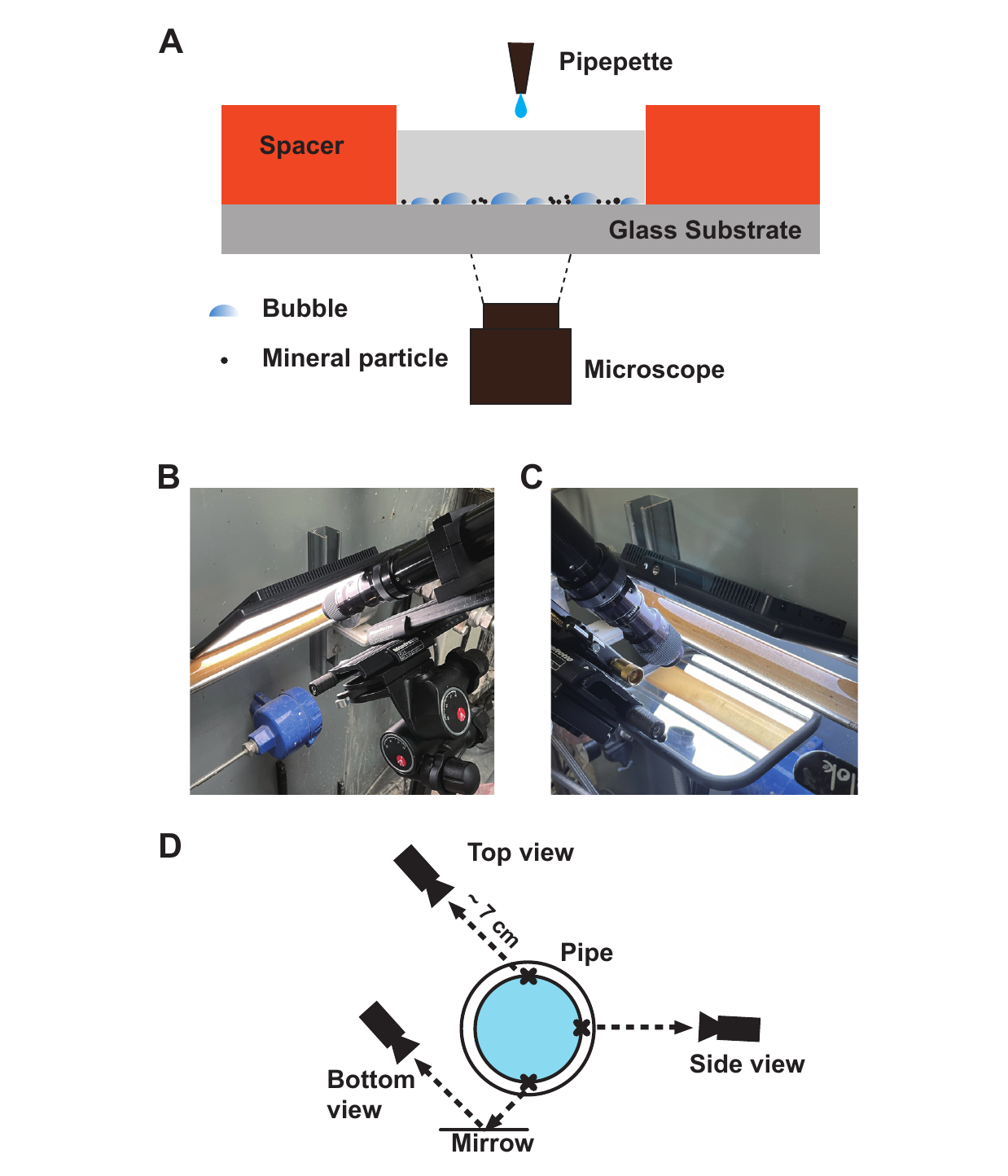}
 \captionsetup{font={normal}}
\renewcommand{\captionfont}{\normalsize}
	\caption{(A) bottom view of bubble generation by $H_{2}O_{2}$ decomposition on the glass substrate. 
 High-speed camera setup: (B) Top positioned lens with a light source, (C) bottom positioned lens with the light source and mirror, and (D) Cross-section sketch of the setup.}
	\label{f02}
\end{figure}

As shown in Figure \ref{f02}A, a modified glass substrate with a spacer was applied for observing the dynamics of bubbles. The microscope with the objective magnification of 10 $\times$ at the bottom was mounted to a Nikon camera having the frame rate of 100 $fps$ presented. In order to investigate the effect of compositions on $H_{2}O_{2}$ decomposition, bitumen and solids were dispersed on the substrate surrounded by the spacer. 500 $\mu$L process water and 20 $\mu$L of 30 wt\% $H_{2}O_{2}$ were injected subsequently. Apart from the room temperature, the temperature was heated to 42 \textdegree C to investigate the effect of thermal energy on the decomposition.

\subsection{High-Speed Camera Setup and Image Analysis}
In Figure \ref{f02}, a Basler high-speed camera was used to capture the images of moving particles in the tailings slurry at the maximum 165 $fps$ for 10 seconds. The camera was mounted with a Navitar's Zoom 6000 Lens System and had $2040 \times 1086$ pixel resolution. By calibration, each pixel is equal to 3.8 $\mu$m. The lens system has a 2$\times$ adapter, allows the dynamic magnification up to 13$\times$. The working distance is 7 $cm$.
The strong light source was installed near the view window for improving the image quality at low exposure time. For comparison, the images were taken from three positions of the cross-section area including top (Figure \ref{f02}B and C), side and bottom view (Figure \ref{f02}D). The mirror image of the bottom was captured from the mirror due to the limitation of space.

Images were further processed by ImageJ software to obtain the size and number distribution of bitumen particles \cite{lind2012open}. Due to the difference in refractive index, the bitumen particles is easy to be distinguished from the background. The image processing consists of several subsequent operations. First, the RGB image was transferred to 8-bit. Second, bandpass filter was applied to remove the noise in the background followed by threshold adjustment. The last operation is the use of particle tracking function for particle counting and measurement.

\section{Results and Discussion}
\subsection{Enhanced Bitumen Recovery with $H_{2}O_{2}$ Addition}

\subsubsection{Effect of $H_{2}O_{2}$ concentration}
\begin{figure}[ht] 
	\centering
	\includegraphics[width=0.8\columnwidth]{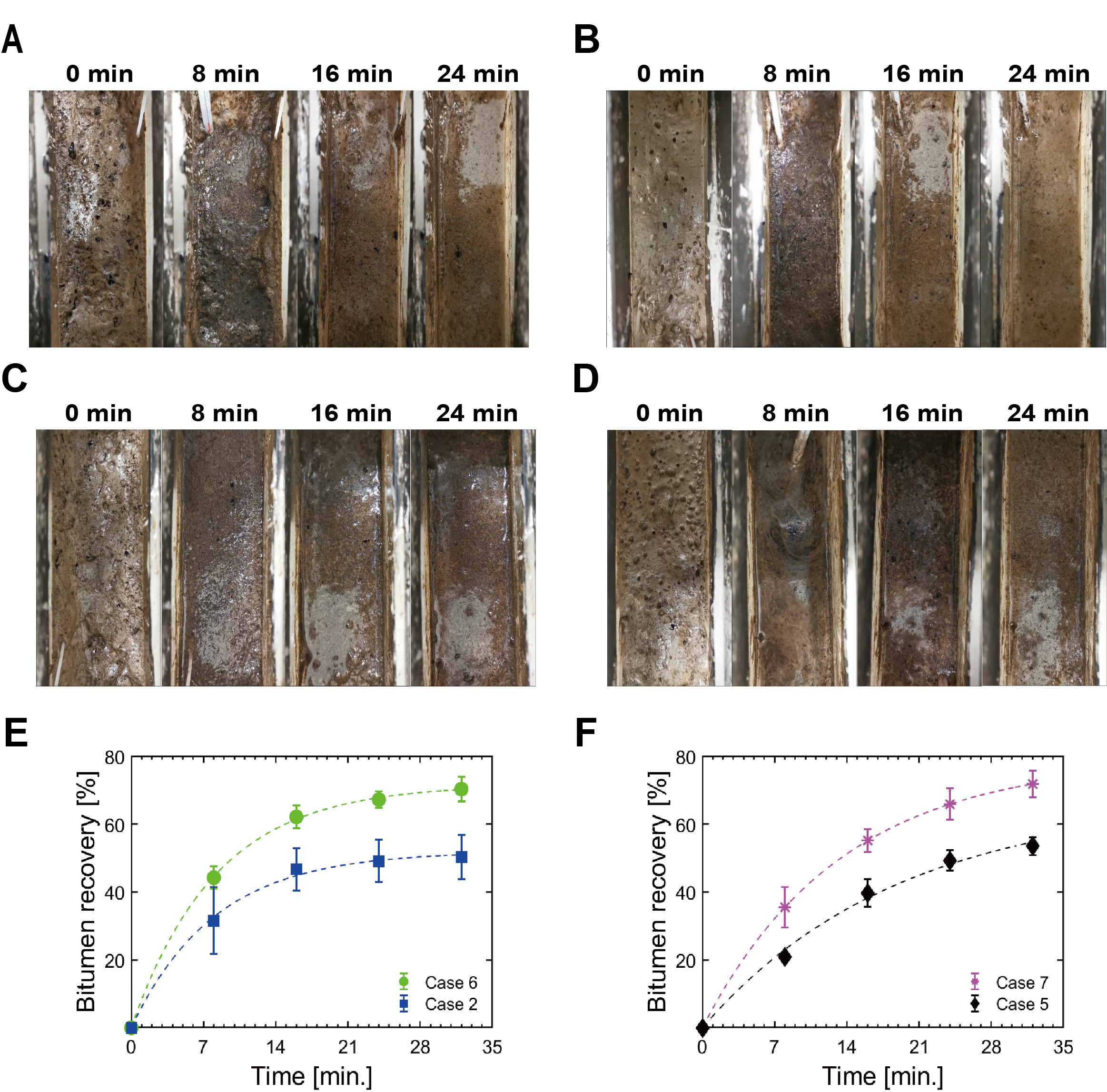}
 \captionsetup{font={normal}}
\renewcommand{\captionfont}{\normalsize}
	\caption{Observation of recovered froth in the trough: (A) Case 6, (B) Case 7, (C) Case 2, and (D) Case 5. Bitumen recovery with injection time: (E) 5 $min$: case 2, 6, and (F) 25 $min$: case 5, 7.}
	\label{f04}
\end{figure}

The compositions of artificial tailings are listed in Table \ref{tab1}. Figure \ref{f04} displays the appearance of recovered froth before each collection and corresponding cumulative bitumen recovery with time. As shown in Figure \ref{f04}A and B,froth in 8 minutes but the amount of recovered froths shrank rapidly at 16 and 24 $min$. It can be inferred that the bitumen recovery drops significantly since 16 $min$. Furthermore, many larger dark bitumen droplets are visible at the surface of collected froth with a higher rate in concentration rate at 0 $min$. The appearance of froth since 16 $min$ with the rate increase of 200 ppm/min is slightly darker than that with the rate increase of 40 ppm/min. In Figure \ref{f04}C and D, the difference in the appearance of recovered froths was less from 0 to 24 $min$ when the injection time was 25 minutes. Additionally, the recovered froths at 16 and 24 $min$ were thicker compared to that in Figure \ref{f04}A and B. In other words, the bitumen recovery rate was more consistent for 32 $min$ with a longer injection duration.

Figure \ref{f04}E shows the bitumen recovery with an injection duration of 5 minutes. For both rates in concentration increase, the bitumen recovery increased to 45\% and 60 \% separately at 16 $min$ but reached plateaus with correspondingly small increases of 8 \% and 10 \% after 16 minutes. The trend of bitumen recovery has a good agreement with the observations as illustrated in Figure \ref{f04}A and B. The rate in concentration rate of 100 ppm/min contributes to the bitumen recovery above 50 \% at 32 $min$, which is about 20 \% lower than the recovery with the rate in concentration increase of 200 ppm/min. It mainly resulted from the difference of 15 \% in bitumen recovery at 8 $min$.
Similarly, as shown in Figure \ref{f04}F, a higher bitumen recovery above 70 \% was obtained with a higher rate in concentration rate of 40 ppm/min. The difference is that the longer injection duration contributes to the continuous bitumen recovery till 32 $min$ which corresponds well with findings in Figure \ref{f04}C and D. 

It is interesting to find that the recovered bitumen at 8 $min$ makes up a large proportion of the cumulative bitumen recovery at 32 $min$. Moreover, the discrepancy of bitumen recovery due to the rate in concentration increase with the same injection duration is caused at 8 $min$. The higher rate of concentration increase, the higher bitumen recovery at 8 $min$ is. At the beginning of injection, it is the initial saturating process of the catalytic sites by $H_{2}O_{2}$ prior to decomposition \cite{lousada2012mechanism}. Consequently, the higher rate in concentration increase enhanced the adsorption of $H_{2}O_{2}$ molecules to catalytic sites for faster and higher production of bubbles which enhances bitumen recovery at the beginning. However, the high concentration of $H_{2}O_{2}$ might have a negative influence that limits decomposition rate of $H_{2}O_{2}$ due to the increased activation energy of decomposition \cite{hiroki2005decomposition}. With continuous injection of $H_{2}O_{2}$, the catalytic surface was supplemented substantially with $H_{2}O_{2}$ and thus the bitumen recovery grew obviously for 32 $min$ in Figure \ref{f04}F.

\begin{figure}[ht] 
	\centering
	\includegraphics[width=0.9\columnwidth]{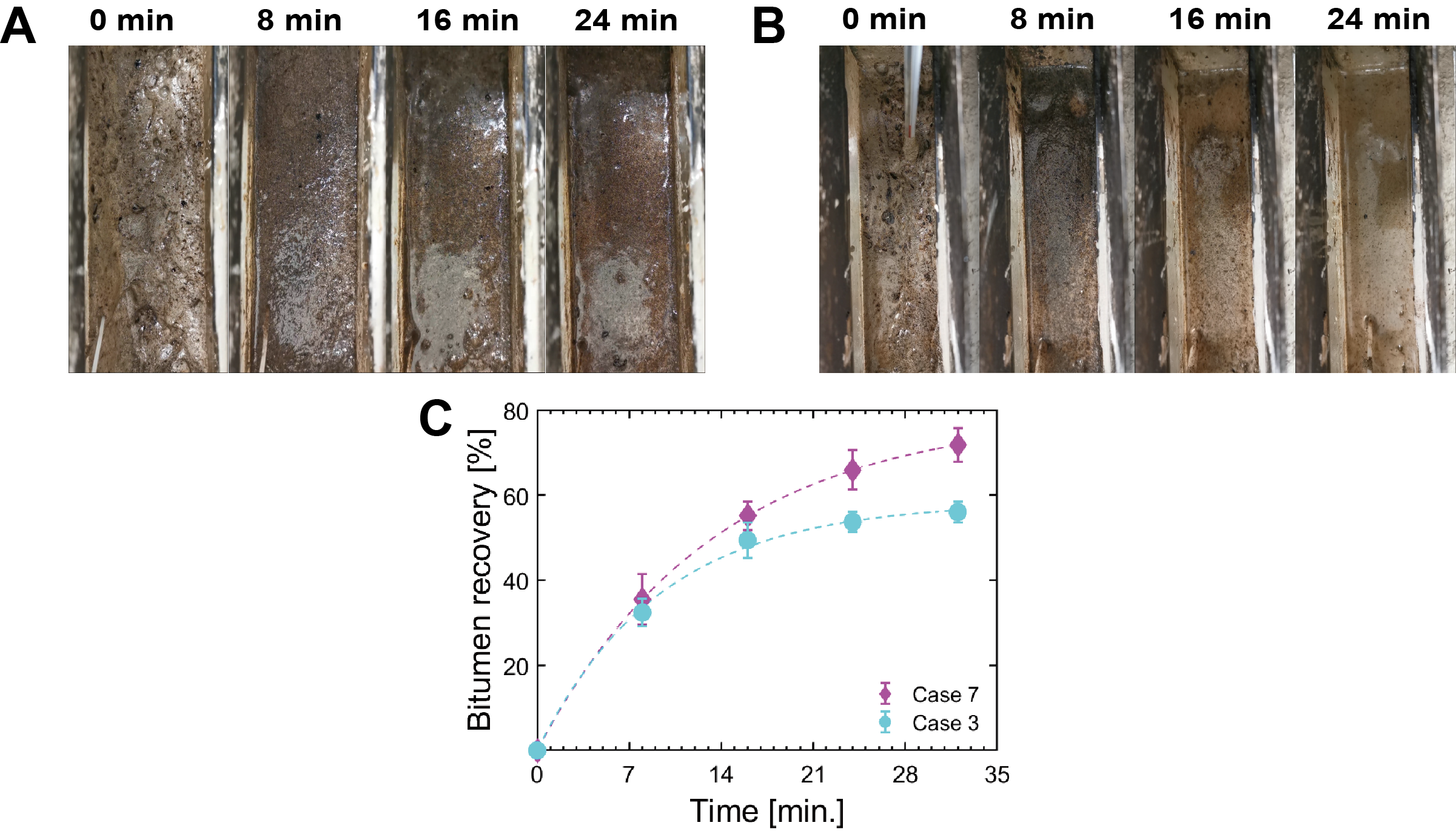}
 \captionsetup{font={normal}}
\renewcommand{\captionfont}{\normalsize}
	\caption{Observation of recovered froth in the trough: (A) 25 $min$, and (B) 12 $min$. Bitumen recovery with the same concentration increase rate: (C) 6 $mL$ $H_{2}O_{2}$ injected for 12 minutes and 12 $mL$ $H_{2}O_{2}$ injected for 25 minutes.}
	\label{f05}
\end{figure}

Figure \ref{f05}A and B show the recovered froths with similar rates in concentration increase and various injection duration. By comparison, the quality of froths recovered with the injection duration of 12 $min$ is worse because fewer dark bitumen droplets were recovered in the froths after 16 $min$ compared to the observation with an injection duration of 25 $min$. The data in Figure \ref{f05}C is consistent with the observation. A similar bitumen recovery of approximately 35 \% was achieved at 8 $min$, but the difference in bitumen recovery was enlarged gradually due to an asymptotic point of bitumen recovery that was already achieved at 16 $min$ with the shorter injection duration.

It is worth noting that bitumen recovery grows significantly within the injection duration in all conditions. It indicates that the generation of bubbles by $H_{2}O_{2}$ decomposition greatly enhances the bitumen recovery efficiency. However, the decomposition of residual $H_{2}O_{2}$ after injection did not facilitate a further improvement in bitumen recovery. As injection of $H_{2}O_{2}$ ends, the concentration decreases with time which possibly reduces the activation energy of decomposition and then accelerate the decomposition of $H_{2}O_{2}$ \cite{hiroki2005decomposition}. At a low concentration, the number of produced bubbles has a sharp decrease although the decomposition rate might be improved. 
Simultaneously, the probability in bubble-bitumen interaction can be decreased with a smaller number of bubbles available.

\subsubsection{Effect of Injection Duration}
\begin{figure}[ht] 
	\centering
	\includegraphics[width=0.8\columnwidth]{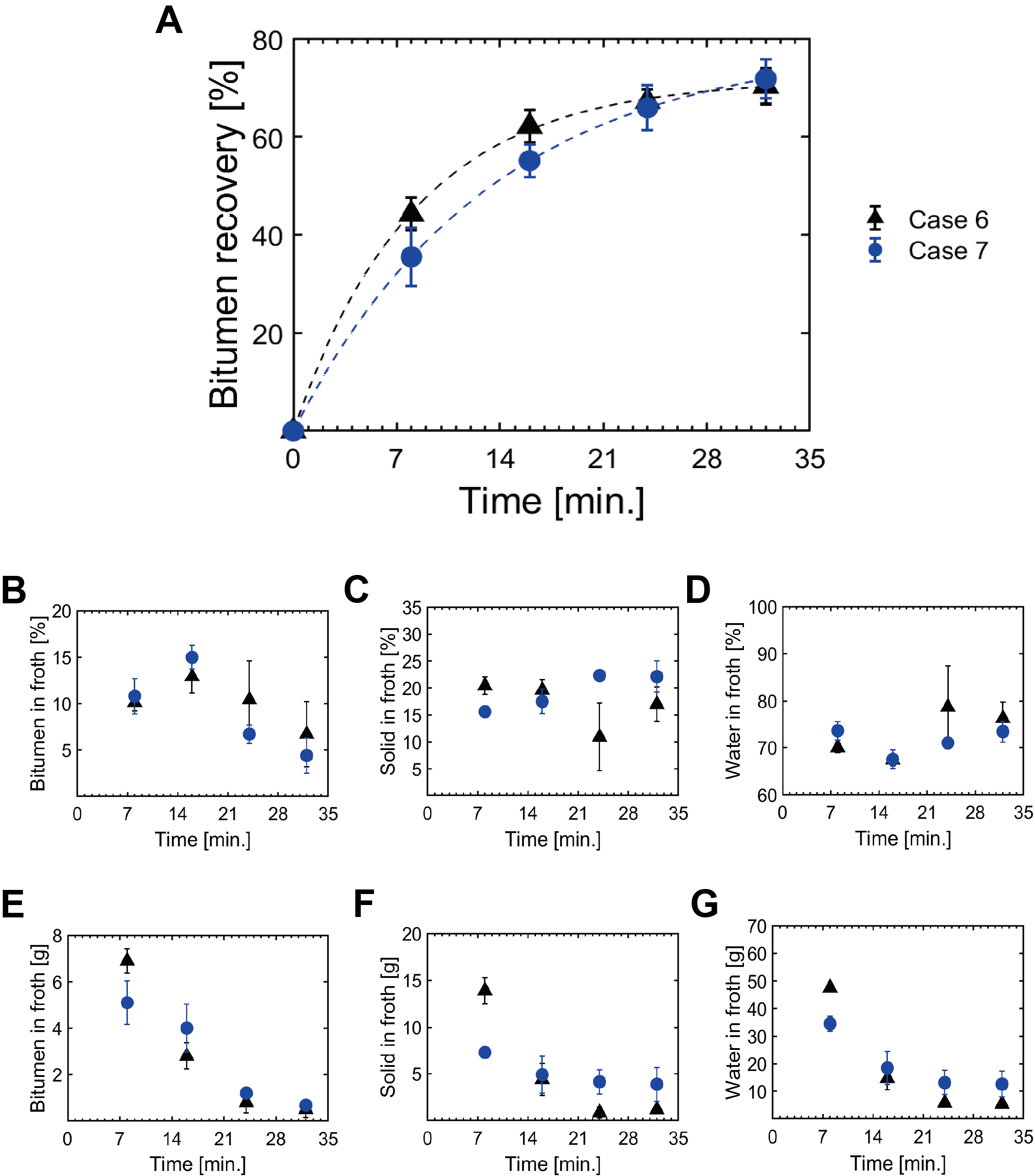}
 \captionsetup{font={normal}}
\renewcommand{\captionfont}{\normalsize}
	\caption{Bitumen recovery with 12 $mL$ $H_{2}O_{2}$: (A) Injection time: 5 $min$ and 25 $min$. Composition in the recovered froth: (B) bitumen, (C) solid, and (D) water. Weight of recovered impurities: (E) bitumen, (F) solid, and (G) water.}
	\label{f06}
\end{figure}

For a constant volume of injected $H_{2}O_{2}$, the injection duration is inversely proportional to the rate of concentration increase. Five various injection duration for 6 $mL$ $H_{2}O_{2}$ and 2 injection duration for 12 $mL$ $H_{2}O_{2}$ are investigated separately. Figure \ref{f06}A displays the bitumen recovery with continuous injection of 12 $mL$ $H_{2}O_{2}$ for 5 and 25 minutes. At first, the bitumen recovery is higher with a duration of 5 minutes. Then, the similar cumulative bitumen recovery of 65 \% and 72 \% was successively achieved at 24 and 32 $min$ with both injection duration. Figure \ref{f06}B-D shows the compositions of bitumen, solid, and water in the froth. The injection duration makes little difference to the compositions, in which bitumen lower than 20 wt \% is minimum and water ranging from 68 \% to 80 \% is maximum. Figure \ref{f06}E-G presents the recovered weight of compositions. The shorter injection duration leads to a much higher recovered weight of all compositions at 8 $min$, but the recovered weight of all compositions drops to a similar or even lower value to that of the longer injection of 25 $min$ since 16 $min$.

\begin{figure}[ht]
	\centering
	\includegraphics[width=0.7\columnwidth]{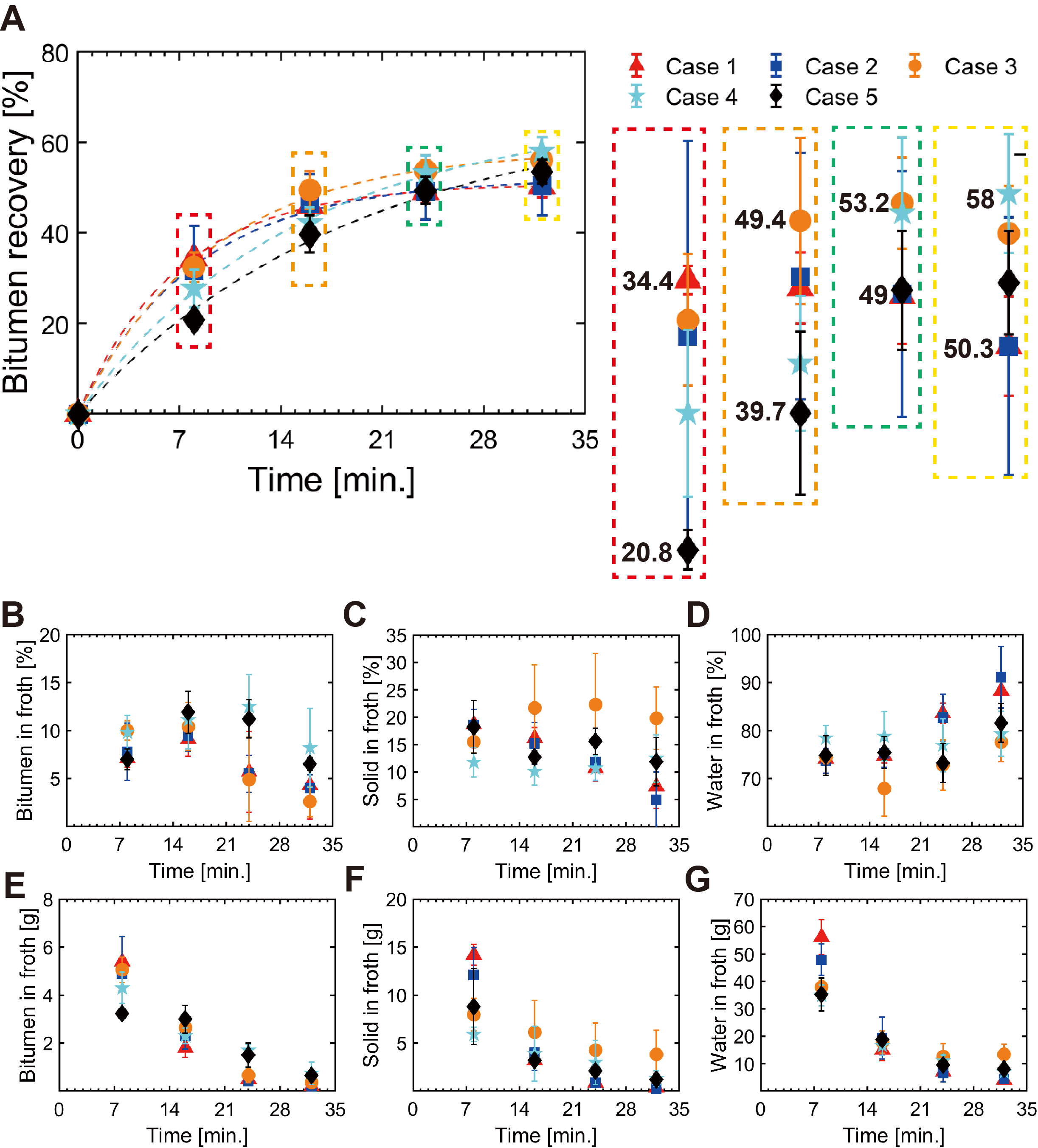}
 \captionsetup{font={normal}}
\renewcommand{\captionfont}{\normalsize}
	\caption{Bitumen recovery with 6 $mL$ $H_{2}O_{2}$: (A) Injection time: 2 $min$, 5 $min$, 12 $min$, 18 $min$, and 25 $min$, the overlapped symbols are scattered on the right. Composition in the recovered froth: (B) bitumen, (C) solid, and (D) water. Weight of recovered impurities: (E) bitumen, (F) solid, and (G) water.}
	\label{f07}
\end{figure}

The cumulative bitumen recovery with 6 $mL$ $H_{2}O_{2}$ injected for various duration is presented in Figure \ref{f07}A. The trend line of bitumen recovery varies with injection duration, similar to the findings in Section 3.1.1. A continuous bitumen recovery is observed with a longer injection duration compared to the negligibly increased recovery efficiency with a shorter injection duration as the injection ends. In addition, cumulative bitumen recovery ranging from 50 to 60 \% at 32 $min$ was achieved with different injection duration. The longer injection duration shows a slightly better performance in enhancing the bitumen recovery at 32 $min$. The distribution of three compositions in the froth with time is presented in Figure \ref{f07}B-D. It is revealed that the compositions of bitumen and solid have an apparent drop with an increase in the composition of water since 24 $min$ with an injection duration shorter than 12 minutes. In terms of the recovered weight of compositions shown in Figure \ref{f07}E-G, it correlates well with the results in Figure \ref{f06} that the shorter injection duration enhances the recovered amount of all compositions, especially solid and water at 8 $min$.
 
Overall, the injection duration of a constant dosage of $H_{2}O_{2}$ only has an impact on the growing trend of bitumen recovery instead of the cumulative bitumen recovery in the end. The shorter injection duration promotes the rate of concentration increase within the process of injection, producing a larger number of bubbles that enhance the bubble-particle collision and attachment. Thus, a greater weight of bitumen and solids along with trapped water is obtained at 8 $min$. However, the number of generated bubbles decreased rapidly after injection, resulting in no further improvement in bitumen recovery. The extension of injection duration decreases the rate of concentration increase but sustains the bubble generation for longer. The continuous growth of bitumen recovery due to a smaller gas production rate for longer can make up for the initial difference in bitumen recovery caused by the higher rate in concentration increase with a shorter injection duration.

\subsubsection{Effect of Additional Catalyst}

Effective catalysts have a great improvement in the rate of $H_{2}O_{2}$ decomposition and bubble formation. Therefore, it is important to understand the effect of amount of catalysts on decomposition. Separated magnetic catalyst of 0.4 g was added to the artificial oil sands tailings by two methods. Two different samples were obtained with the same amount of catalysts added to the bitumen (AC 1) and dry sands (AC 2) separately with subsequent stirring. The operating conditions of case 3 was applied which contributed to a relatively low bitumen recovery of 56 \%. However, no further improvement in bitumen recovery was achieved with both methods of catalyst addition. Although the type of catalysts plays an important role in decomposition, the highly concentrated slurry contains various minerals as shown in Table \ref{tab1} with sufficient catalytic sites for $H_{2}O_{2}$ decomposition. Comparatively, the efficiency of added catalysts is insignificant. Furthermore, the surface of catalysts might be covered by bitumen in the process of mixing. Aggregates of bitumen and catalysts have a higher density than bitumen which might inhibit the flotation of bitumen.

\begin{figure}[ht] 
	\centering
	\includegraphics[width=0.6\columnwidth]{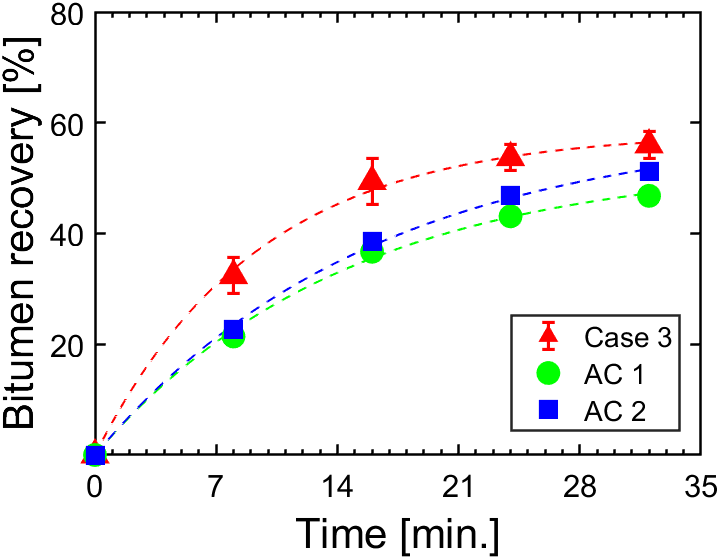}
 \captionsetup{font={normal}}
\renewcommand{\captionfont}{\normalsize}
	\caption{Bitumen recovery with 6 $mL$ $H_{2}O_{2}$ injected for 12 minutes: AC 1 refers to added catalyst in the bitumen and AC 2 refers to added catalyst in the dry sands.}
	\label{f08}
\end{figure}

\subsection{Bitumen and Bubble Distribution Inside the Pipe}
\begin{figure}[!ht] 
	\centering
	\includegraphics[width=0.9\columnwidth]{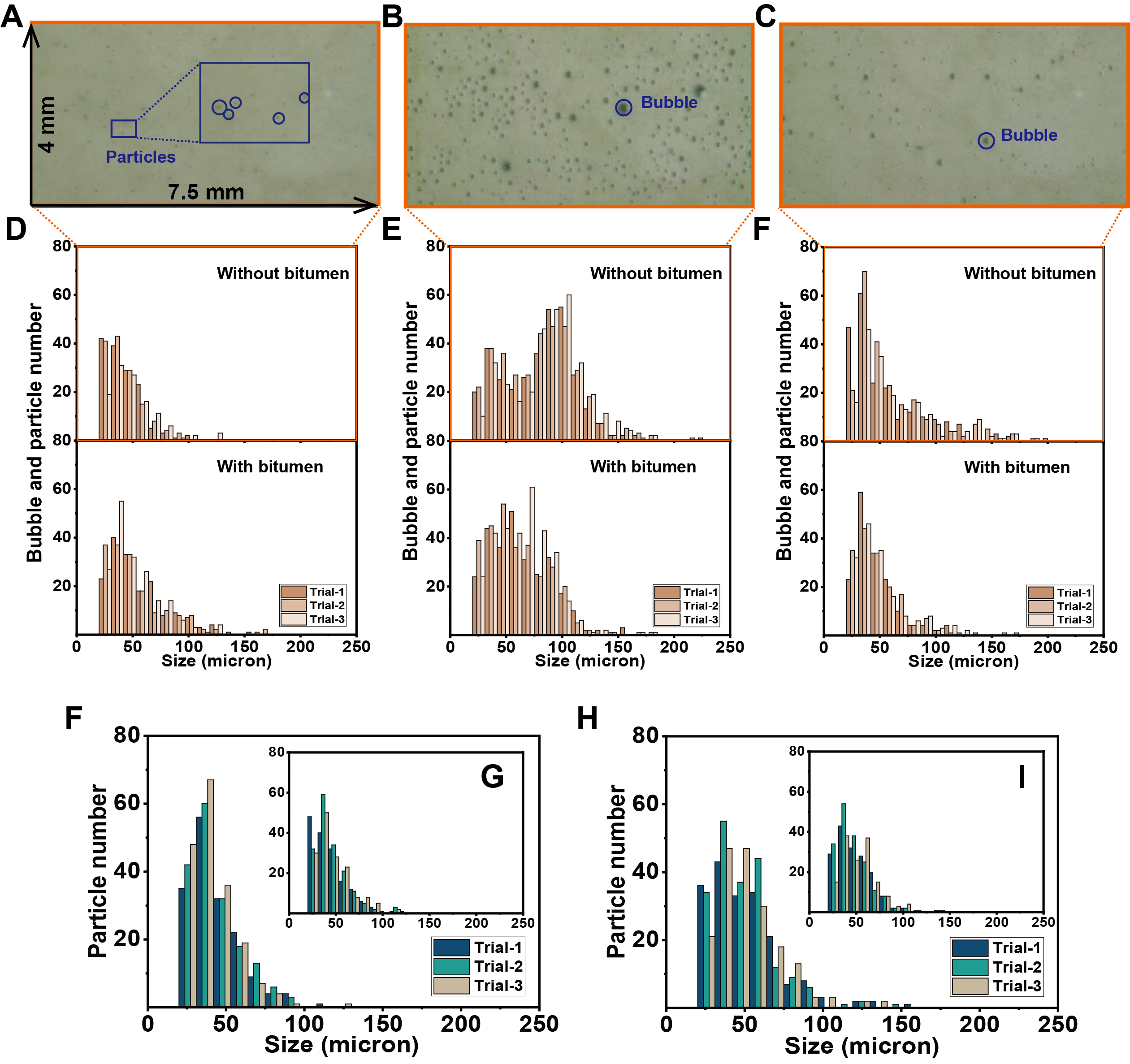}
 \captionsetup{font={normal}}
\renewcommand{\captionfont}{\normalsize}
	\caption{Image analysis of bubbles and particle distribution (Case 4). Screenshots of the top view of bitumen-free tailings with $H_{2}O_{2}$ injection: (A) 0 min, (B) 5 min, (C) 32 min. Size and number distribution of bubbles and particles in the tailings with/without bitumen with $H_{2}O_{2}$ injection: top view: (D) 0 min, (E) 5 min, (F) 32 min; side view: (F) 0 min without bitumen, (G) 5 min with bitumen; bottom view: (H) 0 min without bitumen, (I) 5 min with bitumen.}
	\label{f09}
\end{figure}

The high-speed camera recorded the dynamics of moving particles and bubbles near the pipe wall from top to bottom. 
Figure \ref{f09} shows the distinct observation of bubbles at the top and the size and number distribution of particles and bubbles by image analysis. For comparison, the tailings with and without bitumen were studied separately.
In Figure \ref{f09}A-C, the images with the length of 7.5 $mm$ and width of 4 $mm$ shows the top view with $H_{2}O_{2}$ injection from 0 to 32 $min$. A few tiny particles were dispersed at the top without bubbles before injection. After 5 minutes of $H_{2}O_{2}$ injection, the image was uniformly swarmed with bubbles of various sizes in Figure \ref{f09}B. At 32 min, it is interesting to find that the bubble number decreased significantly compared to that at 5 $min$.

The number and size distribution of the mixture of bubbles and particles in the bitumen-free tailings were presented in Figure \ref{f09}D-F. Figure \ref{f09}D depicts that the number peaks at the range of 20-50 $\mu$m and then decreases with increasing the particle size. Nevertheless, the bimodal distribution of bubbles and particles was seen in Figure \ref{f09}E. Additionally, from the same peak at the range of 20-50 $\mu$m, a larger number of 60 was obtained at the peak of 75-110 $\mu$m, which might result from the generated bubbles in Figure \ref{f09}B. Figure \ref{f09}F exhibits the similar trend of distribution as that in Figure \ref{f09}D. The difference is the larger number of bubble and particle for relative size, especially for the size larger than 75 $\mu$m but the number in this range is smaller than that in Figure \ref{f09}E. The results have a good agreement with the observations. It indicates that the bubbles have a distinguishable distribution different from the particles which concentrate in the range of 70-150 $\mu$m. Moreover, bubbles were not generated immediately after the injection of $H_{2}O_{2}$. The decomposition of $H_{2}O_{2}$ depends on the concentration. Thus, many bubbles were formed at 5 $min$ due to the increase in the concentration. Without the continuous injection after 12 $min$, the residual $H_{2}O_{2}$ was consumed with the decrease in concentration. Only a few bubbles were formed at 32 $min$, resulted from the low concentration of $H_{2}O_{2}$.

The distribution of bubbles and particles in the artificial oil sands tailings of 0.2 wt\% bitumen was displayed in the lower half of Figure \ref{f09}D-F for comparison. At 0 min, the presence of bitumen contributed to a similar trend of distribution with a slightly larger number within the size range of 75-125 $\mu$m. From the discussion above, no bubbles were formed at 0 min. The counted particles with the size of 75-125 $\mu$m were composed of bitumen. The bubble and particle number increased significantly at 5 min but a higher number (20-75 $\mu$m) and a lower number (75-125 $\mu$m) compared to the distribution in the bitumen-free tailings. In the process of circulating the artificial oil sands tailings in the loop, the bitumen was milled into smaller particles by solids of high concentration. Moreover, the oxidation of bitumen consumes $H_{2}O_{2}$ rather than forming bubbles. It could decrease the bubble number and inhibit the faster growth of larger bubbles due to the high concentration and coalescence of small bubbles. In Figure \ref{f09}F, the lower graph also shows a significant decrease in the number compared to 5 minutes under the same condition and a tiny reduction in number compared to that without bitumen at 32 min. Similar to the bitumen-free tailings, residual $H_{2}O_{2}$ were almost decomposed, which led to a sharp decrease in the counted number of bubbles. In terms of bitumen, there should be unrecovered bitumen contained in the tailings. The residual bitumen at 32 min was milled to be quite tiny after 32 minutes and might attach to the surface of the pipeline \cite{sadeghi2022cfd}, which is hard to be captured by bubbles and detect by a high-speed camera.

Furthermore, the top view, and the particle distribution of the side, and the bottom views of the view window are presented in Figure \ref{f09}F-I. It is interesting to see that the presence of residual bitumen and injection of $H_{2}O_{2}$ makes little difference to the particle distribution of the side view as shown in Figure \ref{f09}F and G as well as bottom view in Figure \ref{f09}H and I. In other words, the bubbles did not appear at the side and bottom of the view window at 5 min, although plenty of bubbles were at the top. It agrees with the findings in the simulation work \cite{sontti2022computational,sadeghi2022cfd,sadeghi2022computational}. Additionally, it can be learned that the bitumen distribution was not as obvious as bubbles at three positions. By contrast, the bitumen of a small fraction of 0.2 wt\% had a smaller volume compared to bubbles. Bitumen has a higher density than bubbles, contributing to lower flotation kinetics. Moreover, the tiny bubbles have a lower probability of breaking into smaller bubbles due to the higher internal pressure. But bitumen can be easily sheared into smaller drops that are difficult to recognize.

\subsection{Microscope Analysis of Bubble Generation}
\subsubsection{In Artificial Oil Sands Tailings}

\begin{figure}[ht] 
	\centering
	\includegraphics[width=0.9\columnwidth]{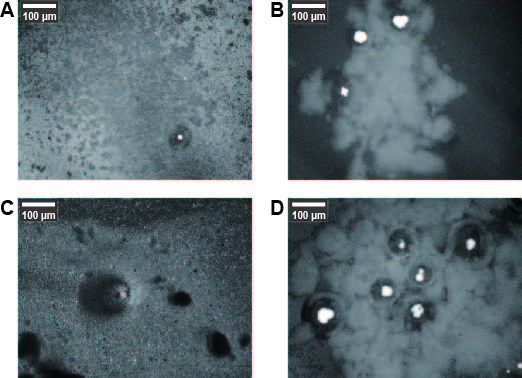}
 \captionsetup{font={normal}}
\renewcommand{\captionfont}{\normalsize}
	\caption{Bubble formation observation with silica: (A) dispersed silica, (B) aggregated silica at room temperature, (C) dispersed silica and (D) aggregated silica at 42 \textdegree C.}
	\label{f11}
\end{figure}

The effect of compositions of the artificial oil sands tailings on bubble formation was examined individually in the 100 ppm $H_{2}O_{2}$ solution. Figure \ref{f11} depicts the bubble formation with the addition of various amounts of non-magnetic particles at room temperature and 42 \textdegree C. According to Table \ref{tab1}, the non-magnetic particles were mostly composed of $SiO_{2}$. Figure \ref{f11}A and C presents that the low efficiency of bubble formation on dispersed non-magnetic particles is independent on the temperature. As the amount of added particles increases, large aggregates formed by tiny particles are shown in Figure \ref{f11}B and D. Notably, a few more bubbles are generated on aggregated particles as shown at both temperatures. The surface of aggregated particles is more rough with numerous holes and crevices compared to the dispersed particles. Generally, the amount and morphology of non-magnetic particles might dominate the bubble formation rather than the temperature.

Those bubbles formed by oxygen molecules are known as the final product of a slow decomposition which is known as a redox reaction \cite{aaaaaaaa}:

\begin{centering}
	
\ch{2 H2O2 -> 2 H2O + O2}

\end{centering}

\noindent Several decomposition mechanisms have been proposed to explain this reaction that might be catalyzed by homogeneous or heterogeneous catalysts. In terms of the nature of the catalyst, the hypothesis applied in the majority of cases is as follows \cite{aaaaaaaa}:

\begin{centering}

\ch{H2O2 -> 2 OH^{.}}

\ch{OH^{.} + H2O2 -> H2O + HO2^{.}}

\ch{OH2^{.} + OH^{.} -> H2O + O2}

\end{centering}
This hypothesis indicates that produced radical intermediates of chain reactions, including the hydroxyl radical (\ch{OH^{.}}) and the peroxy radical (\ch{HO2^{.}}) contribute to the final product (\ch{O2}). 
The alkalinity property of the process water leads to the dissociation of $H_{2}O_{2}$:

\begin{centering}
	
\ch{H2O2 + OH- ->  HO2- + H2O}

\end{centering}

\noindent and thus decreases the amount of $H_{2}O_{2}$. In this case, $H_{2}O_{2}$ acts as a reducing agent in the form of \ch{HO2-} and gives an electron to the metal to generate peroxy radicals (\ch{HO2^{.}})\cite{weiss1935catalytic}. Furthermore, the alkaline solution can accelerate the decomposition of $H_{2}O_{2}$ and oxygen production due to the scavenging of protons as presented in the previous study \cite{molamahmood2022catalyzed}.

Although it is verified that the $SiO_{2}$ surface is able to catalyze the $H_{2}O_{2}$ decomposition according to the detected radicals, the mechanism of the initial catalytic step with the presence of $SiO_{2}$ remains to be studied further \cite{aaaaaaaa}. Specifically, the efficiency of catalysts for $H_{2}O_{2}$ decomposition depends on multi factors, i.e., surface area, material, and temperature. Those factors determine the reaction kinetics according to the standard Arrhenius equation: 
\begin{equation}
	\label{eq:Arrhenius equation}
	ln(k_{obs})= ln(A)- \frac{E_{a}}{RT}
\end{equation}

\noindent where $k_{obs}$ is the observed decomposition rate coefficient, A is the frequency factor, $E_{a}$ is the activation energy, T is the absolute temperature, and R is the gas constant \cite{marr2015high,lousada2012mechanism}. For the specific kind of catalyst, the log of $k_{obs}$ decreases linearly with increasing $\frac{1}{T}$. In other words, high temperatures enhance the rate of reaction due to the improved collision probability in reactive molecules and catalysts \cite{sanchez2011superfast,naeem2020parameters}.
Apart from temperature, the activation energy depending on the catalyst type significantly influences the rate of reaction. \citet{hiroki2005decomposition} found that the lowest decomposition rate was achieved in the presence of $SiO_{2}$ compared to other metal oxides due to the higher activation energy \cite{hiroki2005decomposition}. Moreover, the catalytic surface area, rather than the total surface area, is important to the rate of reaction \cite{lousada2012mechanism}. However, the catalytic surface area might be proportional to the total surface area. As a result, the bubble production is low even at a higher temperature in Figure \ref{f11}C. Comparatively, formed tiny pores and crevices on the particle agglomerates create catalytic sites for $H_{2}O_{2}$ adsorption and restrict the diffusion process, which is beneficial to bubble nucleation and growth \cite{naeem2020parameters}.

\begin{figure}[ht] 
	\centering
	\includegraphics[width=1\columnwidth]{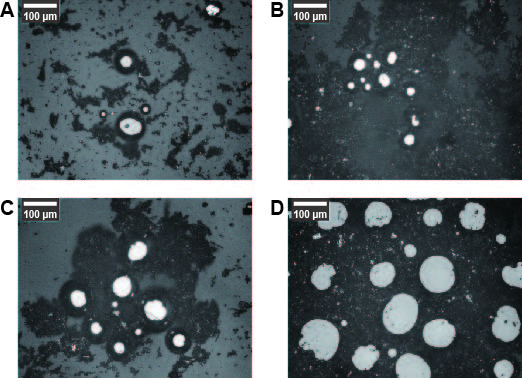}
 \captionsetup{font={normal}}
\renewcommand{\captionfont}{\normalsize}
	\caption{Observation of bubble formation with magnetic particles from sands: (A) low coverage, (B) high coverage at room temperature; (C) low coverage, (D) high coverage at 42 \textdegree C.}
	\label{f12}
\end{figure}

In Figure \ref{f12}, magnetic particles exhibit better performance in generating microbubbles influenced by the temperature and amount of catalysts. Fewer bubbles are visible in the presence of a small amount of magnetic particles at room temperature in Figure \ref{f12}A. A greater number of bubbles were observed on the surface formed by a larger amount of magnetic particles (Figure \ref{f12}B and D). In addition, the higher temperature improved both the bubble generation and growth, as shown in Figure \ref{f12}C and D. As listed in Table \ref{tab1}, the magnetic material was mainly composed of iron oxide with a trace amount of other metal oxides, which can be considered favorable for $H_{2}O_{2}$ decomposition \cite{molamahmood2022catalyzed,ross1958catalytic}. 
Similarly, the higher temperature can greatly increase the probability of a collision between $H_{2}O_{2}$ and catalytic sites.
The process of $H_{2}O_{2}$ decomposition and oxygen production called Fenton reaction in the presence of iron oxide was shown below \cite{molamahmood2022catalyzed,weiss1935catalytic}:

\begin{centering}
	\ch{Fe^{3+}(s) + H2O2 -> ~Fe^{2+}(s) + H+ + HO2^{.}}
	
	\ch{Fe^{2+}(s) + H2O2 -> ~Fe^{3+}(s) + OH- + OH^{.}}
	
	\ch{Fe^{3+}(s) + HO2^{.} -> ~Fe^{2+}(s) + O2 + H+}
	
\end{centering}

Iron oxides composed of $F_{2}O_{3}$ and $F_{3}O_{4}$ can initiate the decomposition of $H_{2}O_{2}$ with the product of both hydroxyl radical (\ch{OH^{.}}) and the peroxy radical (\ch{HO2^{.}}) simultaneously that activate the formation of oxygen. The reduction of $F_{3}O_{4}$ also facilitates oxygen production. The fast decomposition rate of $H_{2}O_{2}$ with iron oxides accounts for the higher oxygen production rate.

Similar to iron oxides, a Fenton-type process is proposed for $H_{2}O_{2}$ decomposition with the presence of $TiO_{2}$ due to the multiple oxidizing states of $Ti^{4+}$ and $Ti^{3+}$ \cite{suh2000reactions}. Those chain reactions can also enhance the decomposition of $H_{2}O_{2}$ further.

\begin{figure}[ht] 
	\centering
	\includegraphics[width=0.8\columnwidth]{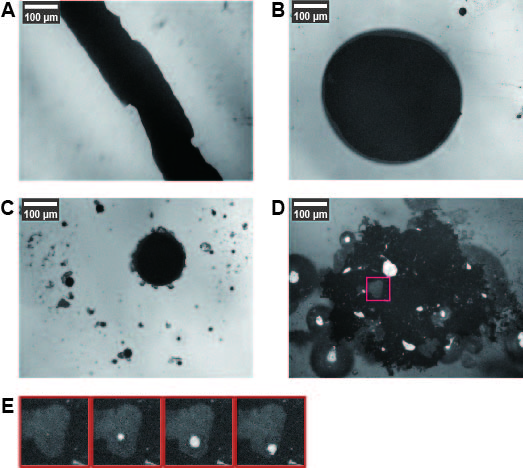}
 \captionsetup{font={normal}}
\renewcommand{\captionfont}{\normalsize}
	\caption{Bubble formation on bitumen: (A) bitumen drop at room temperature; (B) bitumen drop at 42 \textdegree C, (C) bitumen drop mixed with silica at room temperature, (D) bitumen drop mixed with magnetic particles at room temperature, and (E) attachment process of a growing bubble and bitumen surface.}
	\label{f13}
\end{figure}

From the appearance in Figure \ref{f13}A and B, one can see the $H_{2}O_{2}$ decomposition was negligible regardless of the temperature. Thus, bitumen was not a catalytic material for $H_{2}O_{2}$. Similarly, no bubbles appeared with a few non \textendash magnetic particles adsorbed on the bitumen surface (Figure \ref{f13}C). In spite of this, it has been well established that $H_{2}O_{2}$ has been widely used in degrading organic compounds \cite{fernandes2019integrated,boczkaj2017study}. The presence of organic matters suppress the decomposition of $H_{2}O_{2}$ and oxygen production by scavenging radicals at a higher reaction rate for interrupting chain reactions \cite{molamahmood2022catalyzed}. Thus, a competition between oxygen production and degradation of organics would exist with the addition of bitumen. In contrast to the mixture of bitumen and non\textendash magnetic particles, the adsorption of magnetic particles on the bitumen surface contributes to the immediate formation of bubbles on the bitumen surface, as shown in Figure \ref{f13}D. It is found that the generated small bubble attached to the bitumen surface in Figure \ref{f13}E. Apart from the attachment of bubbles to the bitumen, bubbles might be able to nucleate on the bitumen directly due to the embedded catalytic particles. 

\subsubsection{Microbubble Formation in Real Oil Sands Tailings}

\begin{figure}[ht] 
	\centering
	\includegraphics[width=0.9\columnwidth]{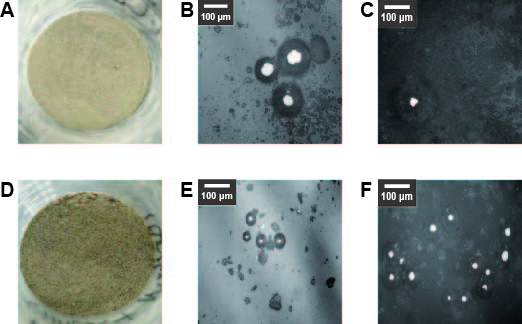}
 \captionsetup{font={normal}}
\renewcommand{\captionfont}{\normalsize}
	\caption{Bubble formation on artificial tailings at room temperature: (A) solid appearance of artificial tailings, (B) low coverage, and (C) high coverage. Real tailings: (D) solid appearance of artificial tailings, (E) low coverage, and (F) high coverage.}
	\label{f14}
\end{figure}

The bubble formation on the sample collected from real oil sands tailings was studied in order to understand  the effect of compositions and the potential of $H_{2}O_{2}$ in the industrial-scale application. Figure \ref{f14} visually compares the appearance of dry samples from artificial and real oil sands tailings and the effect of samples on bubble formation at room temperature separately.
Artificial and real oil sands tailings can be distinguished by the color and size of particles, as illustrated in Figure \ref{f14}A and D. The particles from real oil sands tailings are coarser and darker comparatively. Interestingly, the addition of particles from artificial oil sands tailings did not enhance the bubble formation significantly, as represented in Figure \ref{f14}B. No further improvement in bubble formation was observed with increasing the amount of particles (Figure \ref{f14}C). Compared to Figure \ref{f14}B, the efficiency of bubble formation is similar to the addition of particles from real tailings, as shown in Figure \ref{f14}E. However, the number of bubbles grows rapidly with an increment of the amount of particles in Figure \ref{f14}F.

Oil sand ores have been studied extensively for their solid compositions and subsequently quantified and qualitatively analyzed \cite{osacky2013mineralogical}.
\citet{kaminsky2008characterization} summarized the mineral compositions of oil sand ores and treated froth from a Suncor lease as listed in Table 3 \cite{kaminsky2008characterization}. Quartz is the primary composition in ores and froth, while the fraction decreases in the smaller size ranges. In contrast, smaller particles composed of ilmenite, leucoxene, rutile, and zircon, which contain abundant noble metals such as \ce{Fe},\ce{Ti}, and \ce{Zr}, have a larger fraction in the smaller size ranges. The froth treatment tailings analyzed by \citet{liu2006characterization} demonstrated that the froth treatment tailings contain approximate 10 \% $Fe_{2}O_{3}$, 10 \% $TiO_{2}$ and 3 \% $ZrO_{2}$. Consequently, the difference in the appearance of artificial and real tailings results from the mineral compositions.
Compared to $SiO_{2}$, those oxides are regarded as efficient catalysts for $H_{2}O_{2}$ decomposition dominated by factors including activation energy, number, and efficiency of reactive sites \cite{hiroki2005decomposition}.

Clearly, sufficient real tailings perform better in generating bubbles by decomposition. A large amount of particles provide more surface area and a larger number of reactive sites for $H_{2}O_{2}$. Additionally, more efficient catalysts are contained in a larger sample, especially for real tailings. 

\begin{table}[]
\captionsetup{font={normal}}
\renewcommand{\captionfont}{\normalsize}
\caption{Quantitative composition analysis of coarse froth and ore samples \cite{kaminsky2008characterization}}
\label{tab3}
\begin{tabular}{@{}cccccccccc@{}}
\toprule
\multicolumn{1}{l}{\multirow{2}{*}{}} & \multicolumn{3}{l}{\textbf{Ore}}                                                                                                                                                    & \multicolumn{4}{l}{\textbf{Primary froth}}                                                                                                                                                                                                                & \multicolumn{2}{l}{\textbf{Secondary froth}}                                                                               \\ \cmidrule(l){2-10} 
\multicolumn{1}{l}{}                  & \begin{tabular}[c]{@{}c@{}}\textgreater{}106\\ ($\mu$m)\end{tabular} & \begin{tabular}[c]{@{}c@{}}45-106\\ ($\mu$m)\end{tabular} & \begin{tabular}[c]{@{}c@{}}\textless{}45\\ ($\mu$m)\end{tabular} & \begin{tabular}[c]{@{}c@{}}\textgreater{}45 \\ ($\mu$m)\end{tabular} & \begin{tabular}[c]{@{}c@{}}\textgreater{}106\\ ($\mu$m)\end{tabular} & \begin{tabular}[c]{@{}c@{}}45-106\\ ($\mu$m)\end{tabular} & \begin{tabular}[c]{@{}c@{}}\textless{}45\\ ($\mu$m)\end{tabular} & \begin{tabular}[c]{@{}c@{}}\textgreater{}45\\ ($\mu$m)\end{tabular} & \begin{tabular}[c]{@{}c@{}}\textless{}45\\($\mu$m)\end{tabular} \\ \midrule
\textbf{Quartz}                       & 99$\pm$1                                                             & 95$\pm$1                                                & 43$\pm$2                                                       & 40$\pm$1                                                                  & 97$\pm$1                                                           & 71$\pm$1                                                & 40$\pm$1                                                       & 94$\pm$1                                                          & 34$\pm$2                                                       \\
\textbf{Microline}                    & 1$\pm$1                                                              & 1$\pm$1                                                 & 6$\pm$2                                                        &                                                                       & 1$\pm$1                                                            & 4$\pm$1                                                 & 8$\pm$1                                                        & 2$\pm$1                                                           & 3$\pm$1                                                        \\
\textbf{Plagioclase}                  &                                                                  & Trace                                               & 1$\pm$1                                                        &                                                                       &                                                                & 1$\pm$1                                                 & Trace                                                      &                                                               &                                                            \\
\textbf{Siderite}                     &                                                                  &                                                     & 2$\pm$1                                                        & 9$\pm$1                                                                   &                                                                & 2$\pm$1                                                 & 8$\pm$1                                                        & Trace                                                         & 11$\pm$1                                                       \\
\textbf{Chlorite}                     &                                                                  &                                                     & 3$\pm$2                                                        &                                                                       &                                                                &                                                     & 2$\pm$1                                                        &                                                               & 2$\pm$1                                                        \\
\textbf{Kaolinite}                    &                                                                  & 2$\pm$1                                                 & 30$\pm$3                                                       & 4$\pm$1                                                                   &                                                                & 2$\pm$1                                                 & 13$\pm$2                                                       & Trace                                                         & 31$\pm$3                                                       \\
\textbf{Muscovite}                    &                                                                  & 1$\pm$1                                                 & 13$\pm$1                                                       &                                                                       &                                                                & 2$\pm$1                                                 & 7$\pm$1                                                        & 1$\pm$1                                                           & 5$\pm$1                                                        \\
\textbf{Pyrite}                       &                                                                  &                                                     &                                                            & 5$\pm$1                                                                   &                                                                & 1$\pm$1                                                 & 3$\pm$1                                                        & Trace                                                         & 1$\pm$1                                                        \\
\textbf{Anatase}                      &                                                                  &                                                     & 1$\pm$1                                                        & 3$\pm$1                                                                   & Trace                                                          & 2$\pm$1                                                 & 4$\pm$1                                                        & Trace                                                         & 4$\pm$1                                                        \\
\textbf{Brookite}                     &                                                                  &                                                     &                                                            & 3$\pm$1                                                                   &                                                                & 1$\pm$1                                                 & 3$\pm$1                                                        & Trace                                                         & 1$\pm$1                                                        \\
\textbf{Ilmenite}                     & \multicolumn{1}{l}{}                                             &                                                     &                                                            & 5$\pm$1                                                                   & Trace                                                          & 3$\pm$1                                                 & 2$\pm$1                                                        & Trace                                                         & 1$\pm$1                                                        \\
\textbf{Rutile}                       & \multicolumn{1}{l}{}                                             &                                                     & Trace                                                      & 20$\pm$1                                                                  & Trace                                                          & 4$\pm$1                                                 & 4$\pm$1                                                        & 1$\pm$1                                                           & 3$\pm$1                                                        \\
\textbf{Schorl}                       & \multicolumn{1}{l}{}                                             &                                                     &                                                            & 7$\pm$1                                                                   & 1$\pm$1                                                            & 4$\pm$1                                                 & 3$\pm$1                                                        & 1$\pm$1                                                           & 2$\pm$1                                                        \\
\textbf{Zicron}                       & \multicolumn{1}{l}{}                                             &                                                     &                                                            & 5$\pm$1                                                                   &                                                                & 4$\pm$1                                                 & 3$\pm$1                                                        & Trace                                                         & 2$\pm$1                                                        \\ \bottomrule
\end{tabular}
\end{table}

\subsection{Further discussion on possible impact of H2O2 on recovery of bitumen residue in real processes}

Our above results showed the potential for enhanced recovery of bitumen residue from tailing slurry. Although $H_{2}O_{2}$ is a strong oxidant, widely used in wastewater treatment. For new applications in bitumen residue recovery, many other factors may need to be evaluated in great details.

It will be necessary to evaluate the effects from addition of $H_{2}O_{2}$ on water recycling and any potential impact on other compositions in water and overall water chemistry. In addition, the long-term effects on stability of mature fine tailings after disposal may also be influenced by the catalytic decomposition. 

The added $H_{2}O_{2}$ itself decomposes to pure water and oxygen. The rate of decomposition increases with rise in temperature, concentration, and pH.
In the presence of other chemical species in water, $H_{2}O_{2}$ decomposition can be catalysed by various redox-active ions or compounds. Certain metal ions abundant in tailings, such as $Fe^{2+}$ or $Ti^{3+}$, can cause the decomposition, forming free radicals such as the hydroxyl radical (HO•) and hydroperoxyl (HOO•). 

In terms of using $H_{2}O_{2}$ to recover bitumen from coarse tailings in pipeline transport, another consideration may be that whether the added $H_{2}O_{2}$ could erode the pipeline made of iron. The oxidation/erosion caused by $H_{2}O_{2}$ decomposition should be similar to the level in natural processes, such as from dissolved oxygen molecules or from rains.

In brief, $H_{2}O_{2}$ may be simple and easily implementable for the pipeline transport to recover residual bitumen by in-situ formation of microubbbles. By controlling $H_{2}O_{2}$ concentrations, it may be possible to exploit its simplicity for practical application and to minimize its potential negative impact.

\section{Conclusions}

In this work, we have demonstrated a simple separation approach for enhanced bitumen residue seggregation in highly concentrated slurry flow. 
Microbubbles spontaneously form from in-situ catalytic decomposition of $H_{2}O_{2}$ at low concentration in artificial tailings flow with approximately 50\% solids.  The residue of bitumen in the slurry preferentially accumulates on the top part of the labscale pipeline used in our experiments. The recovery efficiency of the bitumen is significantly enhanced by addition of $H_{2}O_{2}$. The total amount of added $H_{2}O_{2}$ is key, regardless the injection stream is slow and long, or fast and short. The microscopic images confirmed that magnetic particles present the slurries catalyze $H_{2}O_{2}$ decomposition, leading to the microbubble formation. In less than 30 mins, $H_{2}O_{2}$ injection led to a high recovery efficiency of > 70 \% bitumen from a very low initial concentration of 0.2 \%.  More microbubbles form in the real mature fine tailings than in artificial tailings, due to natural presence of catalytic mineral particles. Segregated bitumen in the top of the pipeline may be separated out without processing the entire volume of the slurry to eliminate oily residues and reduce greenhouse gas emission after tailings disposal.

\section*{Credit author statement}
\textbf{Kaiyu Zhou:} Conducting experiment, Methodology, Formal Analysis, Writing – Original Draft, Writing – review \& editing. \textbf{Somasekhara Goud Sontti:} Conducting experiment, Writing – review \& editing. \textbf{Joe Zhou:} Methodology, Technical support. \textbf{Xuehua Zhang:} Conceptualization, Project administration, Writing – review \& editing, Resources, Supervision.

\section*{Declaration of Competing Interests}
The authors declare that they have no known competing financial interests or personal relationships that could have appeared to influence the work reported in this paper.

\section*{Acknowledgement}
This work was supported by the China Scholarship Council (202008180018), the Institute for Oil Sands Innovation (IOSI) (Project Number IOSI 2019-04 (TA)), and the Natural Science and Engineering Research Council of Canada (NSERC)-Alliance Grant. This research was undertaken, in part, thanks to funding from the Canada Research Chairs Program. The authors are grateful for technical support from IOSI labs at the University of Alberta. 

\bibliography{ref}

\begin{thebibliography}{10}
\expandafter\ifx\csname url\endcsname\relax
  \def\url#1{\texttt{#1}}\fi
\expandafter\ifx\csname urlprefix\endcsname\relax\def\urlprefix{URL }\fi
\expandafter\ifx\csname href\endcsname\relax
  \def\href#1#2{#2} \def\path#1{#1}\fi

\bibitem{small2015emissions}
C.~C. Small, S.~Cho, Z.~Hashisho, A.~C. Ulrich, Emissions from oil sands tailings ponds: Review of tailings pond parameters and emission estimates, J. Pet. Sci. Eng. 127 (2015) 490--501.

\bibitem{chalaturnyk2002management}
R.~J. Chalaturnyk, J.~Don~Scott, B.~{\"O}z{\"u}m, Management of oil sands tailings, Pet. Sci. Technol. 20~(9-10) (2002) 1025--1046.

\bibitem{sadeghi2022computational}
M.~Sadeghi, S.~G. Sontti, E.~Zheng, X.~Zhang, Computational fluid dynamics (cfd) simulation of three--phase non--newtonian slurry flows in industrial horizontal pipelines, Chem. Eng. Sci. 270 (2023) 118513.

\bibitem{da2021recovery}
M.~A. da~Silva, F.~L. Motta, J.~B. Soares, Recovery of residual bitumen, dewatering, and consolidation of oil sands tailings with poly (acrylamide-co-lauric acid), Miner. Eng. 174 (2021) 107248.

\bibitem{li2011ionic}
X.~Li, W.~Sun, G.~Wu, L.~He, H.~Li, H.~Sui, Ionic liquid enhanced solvent extraction for bitumen recovery from oil sands, Energy \& Fuels 25~(11) (2011) 5224--5231.

\bibitem{mukhametshina2014evaluation}
A.~Mukhametshina, A.~Morrow, D.~Aleksandrov, B.~Hascakir, Evaluation of four thermal recovery methods for bitumen extraction, in: SPE Western North American and Rocky Mountain Joint Meeting, OnePetro, 2014, pp. SPE--169543.

\bibitem{dai1999particle}
Z.~Dai, D.~Fornasiero, J.~Ralston, Particle--bubble attachment in mineral flotation, J. Colloid Interface Sci. 217~(1) (1999) 70--76.

\bibitem{you2017investigation}
X.~You, L.~Li, J.~Liu, L.~Wu, M.~He, X.~Lyu, Investigation of particle collection and flotation kinetics within the jameson cell downcomer, Powder Technol. 310 (2017) 221--227.

\bibitem{zhou2020role}
J.~Z. Zhou, H.~Li, R.~S. Chow, Q.~Liu, Z.~Xu, J.~Masliyah, Role of mineral flotation technology in improving bitumen extraction from mined athabasca oil sands a ii. flotation hydrodynamics of water-based oil sand extraction, Can. J. Chem. Eng. 98~(1) (2020) 330--352.

\bibitem{wang2020regulation}
H.~Wang, W.~Yang, X.~Yan, L.~Wang, Y.~Wang, H.~Zhang, Regulation of bubble size in flotation: A review, J. Environ. Chem. Eng. 8~(5) (2020) 104070.

\bibitem{tao2005role}
D.~Tao, Role of bubble size in flotation of coarse and fine particles -a review, Sep. Purif. Technol. 39~(4) (2005) 741--760.

\bibitem{zhou2009role}
Z.~Zhou, Z.~Xu, J.~Finch, J.~Masliyah, R.~Chow, On the role of cavitation in particle collection in flotation--a critical review. ii, Miner. Eng. 22~(5) (2009) 419--433.

\bibitem{li2019study}
M.~Li, A.~Bussonni{\`e}re, M.~Bronson, Z.~Xu, Q.~Liu, Study of venturi tube geometry on the hydrodynamic cavitation for the generation of microbubbles, Miner. Eng. 132 (2019) 268--274.

\bibitem{leighton1995bubble}
T.~Leighton, Bubble population phenomena in acoustic cavitation, Ultrason. Sonochem. 2~(2) (1995) S123--S136.

\bibitem{motamed2020microbubble}
A.~Motamed~Dashliborun, J.~Zhou, P.~Esmaeili, X.~Zhang, Microbubble-enhanced recovery of residual bitumen from the tailings of oil sands extraction in a laboratory-scale pipeline, Energy \& Fuels 34~(12) (2020) 16476--16485.
\newblock \href {https://doi.org/https://doi.org/10.1021/acs.energyfuels.0c03000} {\path{doi:https://doi.org/10.1021/acs.energyfuels.0c03000}}.

\bibitem{cai2009dynamic}
J.~Cai, X.~Huai, X.~Li, Dynamic behaviors of cavitation bubble for the steady cavitating flow, J. Therm. Sci. 18~(4) (2009) 338--344.

\bibitem{vlyssides2004bubble}
A.~G. Vlyssides, S.~T. Mai, E.~M.~P. Barampouti, Bubble size distribution formed by depressurizing air-saturated water, Ind. Eng. Chem. Res. 43~(11) (2004) 2775--2780.

\bibitem{zimmerman2008microbubble}
W.~B. Zimmerman, V.~Tesar, S.~Butler, H.~C. Bandulasena, Microbubble generation, Recent Pat. Eng. 2~(1) (2008) 1--8.

\bibitem{caudwell2022protein}
J.~A. Caudwell, J.~M. Tinkler, B.~R. Johnson, K.~J. McDowall, F.~Alsulaimani, C.~Tiede, D.~C. Tomlinson, S.~Freear, W.~B. Turnbull, S.~D. Evans, et~al., Protein-conjugated microbubbles for the selective targeting of s. aureus biofilms, Biofilm 4 (2022) 100074.
\newblock \href {https://doi.org/https://doi.org/10.1016/j.bioflm.2022.100074} {\path{doi:https://doi.org/10.1016/j.bioflm.2022.100074}}.

\bibitem{peyman2012expanding}
S.~A. Peyman, R.~H. Abou-Saleh, J.~R. McLaughlan, N.~Ingram, B.~R. Johnson, K.~Critchley, S.~Freear, J.~A. Evans, A.~F. Markham, P.~L. Coletta, et~al., Expanding 3d geometry for enhanced on-chip microbubble production and single step formation of liposome modified microbubbles, Lab Chip 12~(21) (2012) 4544--4552.

\bibitem{matsuura2019control}
K.~Matsuura, Y.~Yamanishi, C.~Guan, S.~Yanase, Control of hydrogen bubble plume during electrolysis of water, J. Phys. Commun. 3~(3) (2019) 035012.

\bibitem{fu2009bubble}
B.~Fu, C.~Pan, Bubble growth with chemical reactions in microchannels, Int. J. Heat Mass Transf. 52~(3-4) (2009) 767--776.

\bibitem{min2015ph}
K.~H. Min, H.~S. Min, H.~J. Lee, D.~J. Park, J.~Y. Yhee, K.~Kim, I.~C. Kwon, S.~Y. Jeong, O.~F. Silvestre, X.~Chen, et~al., ph-controlled gas-generating mineralized nanoparticles: a theranostic agent for ultrasound imaging and therapy of cancers, ACS nano 9~(1) (2015) 134--145.

\bibitem{molamahmood2022catalyzed}
H.~V. Molamahmood, W.~Geng, Y.~Wei, J.~Miao, S.~Yu, A.~Shahi, C.~Chen, M.~Long, Catalyzed h2o2 decomposition over iron oxides and oxyhydroxides: Insights from oxygen production and organic degradation, Chemosphere 291 (2022) 133037.

\bibitem{huang2013catalytic}
W.~Huang, M.~Manjare, Y.~Zhao, Catalytic nanoshell micromotors, J. Phys. Chem. C 117~(41) (2013) 21590--21596.

\bibitem{nam2009experimental}
Y.~Nam, J.~Wu, G.~Warrier, Y.~S. Ju, Experimental and numerical study of single bubble dynamics on a hydrophobic surface, J. Heat Transf. 131~(12) (2009).

\bibitem{abu2021vapour}
A.~Abu-Bakr, A.~Abu-Nab, Vapour bubble growth within a viscous mixture non-newtonian fluid between two-phase turbulent flow, Int. J. Ambient Energy (2021) 1--8.

\bibitem{okawa2011recovery}
H.~Okawa, T.~Saito, R.~Hosokawa, T.~Nakamura, Y.~Kawamura, S.~Koda, Recovery of bitumen from oil sand by sonication in aqueous hydrogen peroxide, Jpn. J. Appl. Phys. 50~(7S) (2011) 07HE12.

\bibitem{zhou2022microbubble}
K.~Zhou, S.~G. Sontti, J.~Zhou, P.~Esmaeili, X.~Zhang, Microbubble enhanced bitumen separation from tailing slurries with high solid contents, Ind. Eng. Chem. Res. 132 (2022) 268--274.

\bibitem{wallwork2004processibility}
V.~Wallwork, Z.~Xu, J.~Masliyah, Processibility of athabasca oil sand using a laboratory hydro transport extraction system (lhes), Can. J. Chem. Eng. 82~(4) (2004) 687--695.

\bibitem{lind2012open}
R.~Lind, Open source software for image processing and analysis: picture this with ImageJ, Elsevier, 2012.

\bibitem{lousada2012mechanism}
C.~M. Lousada, A.~J. Johansson, T.~Brinck, M.~Jonsson, Mechanism of h2o2 decomposition on transition metal oxide surfaces, J. Phys. Chem. C 116~(17) (2012) 9533--9543.

\bibitem{hiroki2005decomposition}
A.~Hiroki, J.~A. LaVerne, Decomposition of hydrogen peroxide at water- ceramic oxide interfaces, J. Phys. Chem. B 109~(8) (2005) 3364--3370.

\bibitem{sadeghi2022cfd}
M.~Sadeghi, S.~Li, E.~Zheng, S.~G. Sontti, P.~Esmaeili, X.~Zhang, Cfd simulation of turbulent non-newtonian slurry flows in horizontal pipelines, Ind. Eng. Chem. Res. 61~(15) (2022) 5324--5339.

\bibitem{sontti2022computational}
S.~G. Sontti, M.~Sadeghi, K.~Zhou, E.~Zheng, X.~Zhang, Computational fluid dynamics investigation of bitumen residues in oil sands tailings transport in an industrial horizontal pipe, Phys. Fluids 35~(1) (2023) 013340.

\bibitem{aaaaaaaa}
E.~Giamello, B.~Fubini, M.~Volante, D.~Costa, Surface oxygen radicals originating via redox reactions during the mechanical activation of crystalline sio2 in hydrogen peroxide, Colloids Surf. 45 (1990) 155--165.

\bibitem{weiss1935catalytic}
J.~Weiss, The catalytic decomposition of hydrogen peroxide on different metals, Trans. Faraday Soc. (1935).

\bibitem{marr2015high}
K.~M. Marr, B.~Chen, E.~J. Mootz, J.~Geder, M.~Pruessner, B.~J. Melde, R.~R. Vanfleet, I.~L. Medintz, B.~D. Iverson, J.~C. Claussen, High aspect ratio carbon nanotube membranes decorated with pt nanoparticle urchins for micro underwater vehicle propulsion via h2o2 decomposition, ACS nano 9~(8) (2015) 7791--7803.

\bibitem{sanchez2011superfast}
S.~Sanchez, A.~N. Ananth, V.~M. Fomin, M.~Viehrig, O.~G. Schmidt, Superfast motion of catalytic microjet engines at physiological temperature, J. Am. Chem. Soc. 133~(38) (2011) 14860--14863.

\bibitem{naeem2020parameters}
S.~Naeem, F.~Naeem, J.~Zhang, J.~Mujtaba, K.~Xu, G.~Huang, A.~A. Solovev, Y.~Mei, Parameters optimization of catalytic tubular nanomembrane-based oxygen microbubble generator, Micromachines 11~(7) (2020) 643.

\bibitem{ross1958catalytic}
R.~A. Ross, The Catalytic Decomposition of Hydrogen Peroxide Vapour by Oxides and Mixed Oxides, University of Glasgow (United Kingdom), 1958.

\bibitem{suh2000reactions}
M.~Suh, P.~S. Bagus, S.~Pak, M.~P. Rosynek, J.~H. Lunsford, Reactions of hydroxyl radicals on titania, silica, alumina, and gold surfaces, J. Phys. Chem. B. 104~(12) (2000) 2736--2742.

\bibitem{fernandes2019integrated}
A.~Fernandes, M.~G{\k{a}}gol, P.~Mako{\'s}, J.~A. Khan, G.~Boczkaj, Integrated photocatalytic advanced oxidation system (tio2/uv/o3/h2o2) for degradation of volatile organic compounds, Sep. Purif. Technol. 224 (2019) 1--14.

\bibitem{boczkaj2017study}
G.~Boczkaj, A.~Fernandes, P.~Makos, Study of different advanced oxidation processes for wastewater treatment from petroleum bitumen production at basic ph, Ind. Eng. Chem. Res. 56~(31) (2017) 8806--8814.

\bibitem{osacky2013mineralogical}
M.~Osacky, M.~Geramian, D.~G. Ivey, Q.~Liu, T.~H. Etsell, Mineralogical and chemical composition of petrologic end members of alberta oil sands, Fuel 113 (2013) 148--157.

\bibitem{kaminsky2008characterization}
H.~A. Kaminsky, T.~H. Etsell, D.~G. Ivey, O.~Omotoso, Characterization of heavy minerals in the athabasca oil sands, Miner. Eng. 21~(4) (2008) 264--271.

\bibitem{liu2006characterization}
Q.~Liu, Z.~Cui, T.~Etsell, Characterization of athabasca oil sands froth treatment tailings for heavy mineral recovery, Fuel 85~(5-6) (2006) 807--814.

\end{thebibliography}

\end{doublespace}
\end{document}